\documentclass[a4paper]{article}
\usepackage{arxiv}
\usepackage[utf8]{inputenc}
\usepackage[square,numbers]{natbib}
\usepackage[T1]{fontenc}
\usepackage[labelfont=bf,tableposition=top]{caption}
\usepackage{array,booktabs,tabularx,threeparttable}
\usepackage{float,placeins}
\usepackage{nicefrac}       
\usepackage{microtype} 
\usepackage{doi}
\usepackage{booktabs}       
\usepackage{amsfonts,amsmath,amssymb}       
\usepackage{xspace}
\usepackage{lipsum,amsmath,amssymb,amsfonts}
\usepackage{graphicx}
\graphicspath{ {./images/} }
\usepackage{hyperref}  
\title{A comprehensive analysis of PINNs: Variants, Applications, and Challenges }

\author{
 Afila Ajithkumar Sophiya \\
  Lincoln AI Lab\\
  School of Engineering and Physical Science\\
  University of Lincoln\\
  Lincoln, United Kingdom\\
  \texttt{29511866@students.lincoln.ac.uk} \\
   \And
 Akarsh K Nair \\
  Department of Computer Science \& Engineering\\
  Indian
Institute of Information Technology Kottayam\\
  Kerala, India \\
  \texttt{0000-0002-7734-0367} \\
  \And
 Sepehr Maleki \\
   Lincoln AI Lab\\
  School of Engineering and Physical Science\\
  University of Lincoln\\
  Lincoln, United Kingdom\\
  \texttt{smaleki@lincoln.ac.uk} \\
   \And
 Senthil K. Krishnababu \\
  Siemens Energy\\
  Lincoln, United Kingdom\\
  \texttt{senthil.krishnababu@siemens-energy.com} \\
}

\date{}

\renewcommand{\shorttitle}{A comprehensive analysis of PINNs}
\begin{document}

\maketitle

\begin{abstract}
	Physics Informed Neural Networks (PINNs) have been emerging as a powerful computational tool for solving differential equations. However, the applicability of these models is still in its initial stages and requires more standardization to gain wider popularity. Through this survey, we present a comprehensive overview of PINNs approaches exploring various aspects related to their architecture, variants, areas of application, real-world use cases, challenges, and so on. Even though existing surveys can be identified, they fail to provide a comprehensive view as they primarily focus on either different application scenarios or limit their study to a superficial level. This survey attempts to bridge the gap in the existing literature by presenting a detailed analysis of all these factors combined with recent advancements and state-of-the-art research in PINNs. Additionally, we discuss prevalent challenges in PINNs implementation and present some of the future research directions as well. The overall contributions of the survey can be summarised into three sections: A detailed overview of PINNs architecture and variants, a performance analysis of PINNs on different equations and application domains highlighting their features. Finally, we present a detailed discussion of current issues and future research directions. 
\end{abstract}
\keywords{Physics informed ML\and PINNs\and Deep learning\and Differential Equations.}

\section{Introduction}
Over the years, the advancements in deep learning techniques for solving real-world problems have made the technology gain the attention of both industry and academia. From trivial problems to high-end applications, deep learning (DL) holds the key to effective solution generation. The higher the complexity of the problem, the deeper the architecture of the model required for problem-solving. Over time, deep neural networks (DNN) have already proved their mettle in various tasks associated with computer vision, natural language processing, robotics, and so on. Classical AI tasks such as pattern recognition, event classification, and signal processing which were previously performed by machine learning (ML) algorithms have been revolutionized with the advent of deep learning.

In the current scenario, DL models have found increasing applicability for solving both classical and applied mathematics and physics-related problems as well. When it comes to physics-related problems, developing solutions often require data-driven methodologies. However, the core physics mechanisms in most cases cannot be interpreted well or numerical derivations performed on a resource-constrained scenario. Similarly, with mathematical applications such as solving partial differential equations (PDE), standard numerical approaches tend to be insufficient due to their high complexity. Thus, harnessing its capabilities of performing universal approximation and interpretable result generation,  DL models have tapped into the opportunity as an efficient alternative method for solving such scientific computing problems. Recent research has also showcased the effectiveness of DL for generating meta-models with faster output generation capabilities, especially for spatio-temporal data-based domains. Particularly, neural networks (NN) have been able to identify and represent the hidden relationships among input-output variables in complex use cases. However, such complex systems usually process high dimensional data which then introduces the curse of dimensionality to these systems as well \cite{bellman1966dynamic}. In spite of these, ML and DL-based algorithms are promising prospects for working with PDEs \cite{blechschmidt2021three}.

Physics Informed Neural Networks (PINNs) are a new variant of classical neural networks specifically developed for solving partial differential equations and their different variants. PINNs leverage the approximation capability of NNs and transform a PDE into an unconstrained optimization problem. The loss function is designed such that it reflects various factors associated with the particular PDE such as initial condition, boundary condition, and collocation points of the given PDE. In contrast to regular DL approaches, PINNs accept individual points inside the integration domain as input and generate an estimated solution based on that point for a specific PDE. Thus, integrating a residual network capable of encoding physical equations was the major novelty of PINNs which enabled it to perform such mathematical operations.
Traditional numerical methods divide the domain into grid points based on which solutions are approximated. However, PINNs employ mesh-free methodology thus capable of handling diverse problems with irregular, complex, or high-dimensional geometries where grid discretization tends to be notoriously challenging, which is an added advantage of the model.

The idea of utilizing neural networks to solve differential equations was first proposed by Lagaris et al.~\cite{lagaris1998artificial}. The authors formulated a novel trial solution that included two parts: one that satisfied the initial or boundary condition and the other that satisfied the NN. This arrangement is still regarded as a foundation for addressing numerous differential equations. However, a fully functional proposal was presented by Raissi et al. ~\cite{raissi2019physics} which then triggered extensive research interest in the area. Even though PINNs are one of the most popular methodologies for solving PDE currently, several other approaches are employed as well. The prime focus of the study is to comprehensively present the existing literature on PINNs based on aspects related to working methodology, application domains, variants, and so on. In Table~\ref{table1}, we attempt to provide a short overview of the counterparts and predecessors of PINNs prior to moving into a detailed study on the topic. 

\begin{center}
\begin{table}[!hbt] 
 \caption{A brief summary of similar models and their comparisons with PINNs}\label{table1}
\centering
\scriptsize
\begin{tabular}{|c|l|l|} 

\hline
Existing Models & \multicolumn{1}{c|}{Primary Features} & \multicolumn{1}{c|}{Comparison with PINNs}   \\                                         \hline

\begin{tabular}[c]{@{}c@{}}Multilayer Perceptron\\(MLP)\end{tabular}  &  \begin{tabular}{@{\textbullet \hspace{\dimexpr\labelsep+0.5\tabcolsep}}l@{}}Function approximation capability\\Versatility\\Customizable architecture\\\begin{tabular}[c]{@{}l@{}}Solution is differentiable and \\ of closed analytic form\end{tabular}\\Generalization capability\end{tabular}                                          & \begin{tabular}{@{\textbullet\hspace{\dimexpr\labelsep+0.5\tabcolsep}}l@{}}\begin{tabular}[c]{@{}l@{}} General purpose network used for\\ different ML task\end{tabular}\\\begin{tabular}[c]{@{}l@{}}Large amount of labeled data needed \\for complex problems\end{tabular}\\Computational cost\end{tabular}                                      \\ 
\hline
\begin{tabular}[c]{@{}c@{}}Radial Basis Function Neural Network\\(RBFN)\end{tabular} & \begin{tabular}{@{\textbullet\hspace{\dimexpr\labelsep+0.5\tabcolsep}}l@{}}\begin{tabular}[c]{@{}l@{}}Consists of three layers: input layer, hidden \\layer with radial basis functions and output layer\end{tabular}\\Accurate solution\\Generalization capability\end{tabular}                                                    & \begin{tabular}{@{\textbullet\hspace{\dimexpr\labelsep+0.5\tabcolsep}}l@{}}\begin{tabular}[c]{@{}l@{}}Used for data-driven function\\ approximation and pattern recognition\end{tabular}\\Effective for solving regression problems\end{tabular}  \\ 
\hline
\begin{tabular}[c]{@{}c@{}}Cellular Neural Networks \end{tabular}              & \begin{tabular}{@{\textbullet\hspace{\dimexpr\labelsep+0.5\tabcolsep}}l@{}}Efficient computation\\Higher Accuracy\\Solves complex non-linear and stiff DE\end{tabular}                                                                                                                                                                  & \begin{tabular}{@{\textbullet\hspace{\dimexpr\labelsep+0.5\tabcolsep}}l@{}}\begin{tabular}[c]{@{}l@{}}Spatial discretization into grid points\end{tabular}\\\begin{tabular}[c]{@{}l@{}}Allows parallel computing but communication\\ restricted to neighbouring units only\end{tabular}\\\end{tabular}                                                                                  \\ 
\hline
\begin{tabular}[c]{@{}c@{}}Finite Element Neural Networks\\(FENN)\end{tabular}       & \begin{tabular}{@{\textbullet\hspace{\dimexpr\labelsep+0.5\tabcolsep}}l@{}}Solves forward and inverse problems\\\begin{tabular}[c]{@{}l@{}}Represent finite element mode in parallel form\end{tabular}\\Reduce computational effort\end{tabular}                                                                                  &                                      \begin{tabular}{@{\textbullet\hspace{\dimexpr\labelsep+0.5\tabcolsep}}l@{}}\begin{tabular}[c]{@{}l@{}}Minimal training as weights are ~\\calculated and stored\\\end{tabular}\\ \begin{tabular}[c]{@{}l@{}}No weight updations\end{tabular}  
\end{tabular} \\ 
\hline
\begin{tabular}[c]{@{}c@{}}Deep Galerkin Methods\\(DGM)\end{tabular}       & \begin{tabular}{@{\textbullet\hspace{\dimexpr\labelsep+0.5\tabcolsep}}l@{}}Mainly developed for PDEs\\\begin{tabular}[c]{@{}l@{}}Suited for complex geometrical problems as \\ messless approach is employed \end{tabular}\end{tabular}                                                                                  &                                      \begin{tabular}{@{\textbullet\hspace{\dimexpr\labelsep+0.5\tabcolsep}}l@{}}\begin{tabular}[c]{@{}l@{}}Face convergence issues due to \\constraint evaluation at training time\\\end{tabular}\\ \begin{tabular}[c]{@{}l@{}}Not suited for high dimensional PDE\end{tabular}  
\end{tabular} \\ 
\hline

\begin{tabular}[c]{@{}c@{}}Neural differential equations\\(NDE)\end{tabular}       & \begin{tabular}{@{\textbullet\hspace{\dimexpr\labelsep+0.5\tabcolsep}}l@{}}\begin{tabular}[c]{@{}l@{}}Generate a continuous representation \\of solution for PDE\end{tabular}\\\begin{tabular}[c]{@{}l@{}}Easily integrate physical knowledge into\\ PDE based functions\end{tabular}\end{tabular}                                                                                  &                                      \begin{tabular}{@{\textbullet\hspace{\dimexpr\labelsep+0.5\tabcolsep}}l@{}}\begin{tabular}[c]{@{}l@{}}Computationally expensive for \\high dimensional problems\\\end{tabular}\\ \begin{tabular}[c]{@{}l@{}}Faces convergence issues\end{tabular}  
\end{tabular} \\ 
\hline

\begin{tabular}[c]{@{}c@{}}Graph neural networks\\(GNN)\end{tabular}       & \begin{tabular}{@{\textbullet\hspace{\dimexpr\labelsep+0.5\tabcolsep}}l@{}}\begin{tabular}[c]{@{}l@{}}Applicable to PDEs following \\graph-structured data representations.\\\end{tabular}\\\begin{tabular}[c]{@{}l@{}}Physical components can\\ be integrated\end{tabular}\end{tabular}                                                                                  &                                      \begin{tabular}{@{\textbullet\hspace{\dimexpr\labelsep+0.5\tabcolsep}}l@{}} \begin{tabular}[c]{@{}l@{}}Higher volumes of training data\\is required.\\\end{tabular}\\
\begin{tabular}[c]{@{}l@{}}Computational complexity increases with\\increase in data size\\\end{tabular}\\ \begin{tabular}[c]{@{}l@{}}Lack of generalizability \end{tabular}
\end{tabular} \\ 
\hline

\begin{tabular}[c]{@{}c@{}}Wavelet Neural Networks\\(WNN)\end{tabular}               & \begin{tabular}{@{\textbullet\hspace{\dimexpr\labelsep+0.5\tabcolsep}}l@{}}Tool for~non linear approximation\\ \begin{tabular}[c]{@{}l@{}} Feed-forward NN with input, hidden \& output \\layer  with hidden layer employing wavelet\\ activation function \end{tabular}\\\begin{tabular}[c]{@{}l@{}} Employs translation and dilation parameters
\end{tabular}\\\begin{tabular}[c]{@{}l@{}}High precision during learning and~prediction\\process\end{tabular}\end{tabular} &                                                         \begin{tabular}{@{\textbullet\hspace{\dimexpr\labelsep+0.5\tabcolsep}}l@{}}\begin{tabular}[c]{@{}l@{}}Suitable for time-series data\end{tabular}\\ \begin{tabular}[c]{@{}l@{}}Large amount of training data\end{tabular} \end{tabular}    \\                                  \hline
\end{tabular}
\end{table}
\end{center}

\section{Contributions}
Even though several surveys on PINNs are available such as \cite{cuomo2022scientific}, they only provide a birdseye
view of the technology and fail to deliver an extensive analysis. The existing surveys have been disintegrated with most of the articles either having minimal literature on PINNs~\cite{bdcc6040140} or being solely limited to general applications~\cite{en16052343,9743327}. Thus we attempt to bridge the gap in the existing literature by providing an in-depth analysis of different aspects of the technology such as technological concepts, modes of applications,
various use cases, and so on.

The contributions of the survey can be summarised as follows :
\begin{enumerate}
    \item \textbf{Historic Analysis :} A short study on the predecessors and counterparts of PINNs is presented. Various frameworks were evaluated and their features were compared and contrasted against PINNs to provide an empirical understanding of the topic.
    \item \textbf{Comprehensive Analysis :} The study presents a detailed overview of the most recent state-of-the-art PINN models from various areas with a special notion of their architecture, methodology, and other features.
    
    \item \textbf{Architectural Overview:} The general architecture of PINNs models with a comprehensive discussion on individual components along with mathematical formulation is presented.
    \item \textbf{Listing down major variants :} An exclusive study focusing on identifying and differentiating variants of PINNs has been performed. Even though several other variants could be identified, only articles with commendable research contributions have been included in this study.

    \item \textbf{Evaluating solvability :} A detailed study of PINN-based approaches for solving different classes of equations is presented, namely ordinary differential equations, partial differential equations, and fractional differential equations.
\item \textbf{Real-world applications :} The article presents comprehensive case studies assessing the applicability of
PINN models on various real-life applications and their limitations. 

\item \textbf{ Limitations and Futurscope :} Consolidating from various studies, we list down some of the major problems faced in the implementation of the PINN models. We also highlight prospective research gaps for further studies on the topic.

\end{enumerate}

The rest of this study is structured into eight sections as follows. In Section~\ref{sec1}, we present the basic PINN architecture and explain some of the major terminologies associated with the technology. In Section~\ref{sec2} and ~\ref{sec3}, a comprehensive study of PINNs for solving ODE and PDE is discussed respectively. Following that, in Section~\ref{sec4}, we present the existing literature on fPINNs, an exclusive PINNs model for solving fractional PDEs. In section~\ref{sec5}, we list down some of the most popular variants of PINNs followed by major applications of PINNs in real-world scenarios in section~\ref{sec6}. Section~\ref{sec7} presents some of the common challenges associated with PINNs implementation and Section~\ref{sec8} concludes the study.

\section{PINNs Architecture}\label{sec1}

Physics-informed Neural Networks are a variant of the classical neural networks that incorporate scientific ML approaches for solving problems related to differential equations (DE). Rather than directly solving DEs, PINNs approximate the solutions from the knowledge gained via neural networks. These approximations are further optimized by minimizing the divergence between the values generated and the expected values with the aid of a specified loss function. The function is influenced by several parameters specific to the equation being solved such as initial and boundary conditions, spatial and temporal domain as well as the residual points of the DE usually referred to as collocation points. Thus, PINNs can be considered regular DL models with the capability to estimate solutions for DEs when provided with input points in the specified integration domain trained for. However, the novelty of the system lies in the ability of the NN employed to understand and encode the physical knowledge required for generating the approximations. Even though not synonymous, the concept of PINNs training can be equated to the classical unsupervised learning strategy where the models are capable of deriving inferences from unlabeled training data. In contrast to the majority of deep learning approaches, PINNs employ a mesh-free approach for identifying solutions to DE. Rather than solving the equation in a single-shot manner, the models initially convert the equations into loss function optimization problems and then solve them. For this, dedicated mathematical models are utilized and the loss function is reinforced using penalties from the residual term obtained from the preset governing equations. This also creates a restricting effect to the model thus limiting it to identifying solutions inside the acceptable boundaries only.

\begin{figure*}
\centering
\includegraphics[width=15cm]{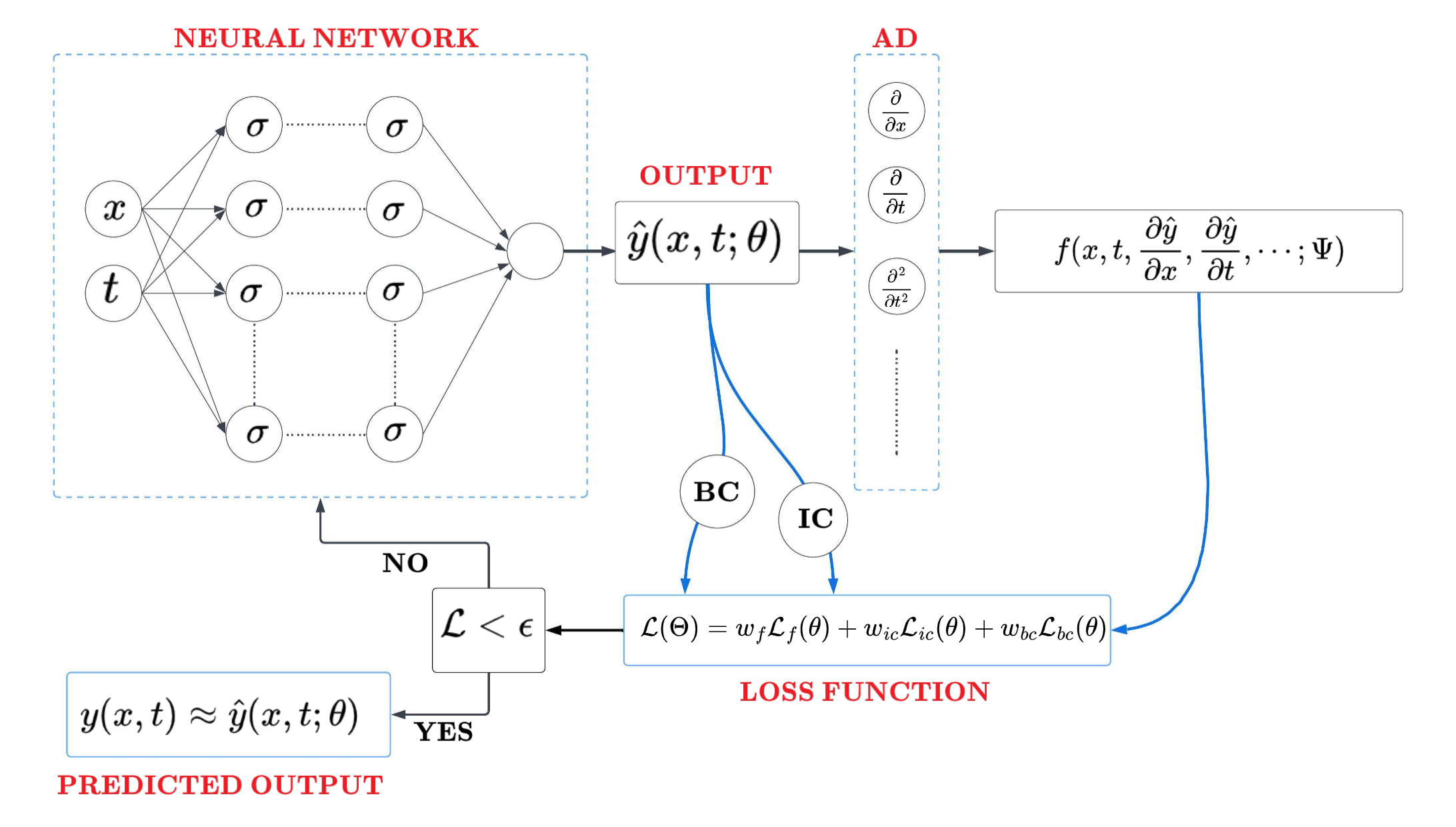}
\caption{Generalised architecture of PINNs model }
\label{figure1}
\end{figure*}

 In this section, we attempt to outline some of the major components of a PINN model. Even though the majority of the techniques employed in a PINN network such as loss function, physical knowledge integration, and so on are model-centric procedures, we present the underlying principles behind the formulation of such components. Figure.~\ref{figure1} depicts a basic PINN model architecture outlining the placement of the various components. In general, PINNs architecture comprises three components, a neural network, an automatic differentiation module, and a loss function. Each of these components needs to be designed for the specific use case to generate optimal results. The respective components and their working methodologies are explained as follows :

\subsection{Deep Neural Networks}
Deep Neural Network  is a subset of ANN employed to model and resolve complex problems. In general scenarios, a significant amount of data is needed to train the NN to capacitate it to recognize patterns and make appropriate predictions or decisions without explicit programming. DNNs have the ability to automatically learn from experience and improve over time, increasing their accuracy. 

Consider a function $G^H(x):\mathbb{R}^{N_1}\longrightarrow\mathbb{R}^{N_H}$ be a $H$ layer neural network with $N_h$ neurons in $h^{th}$ layer. The layer between the input layer $N_1$ and the output layer $N_H$ is the hidden layer. Let $W_h\in \mathbb{R}^{N_h}\times\mathbb{R}^{N_{h-1}}$ be the weight matrix in the $h^{th}$ layers, which can be taken as an indicator of a neuron's sensitivity to each input term. Then, $W_h$ can be represented as : 
\begin{equation*}
W_h=
  \begin{bmatrix}
    w_{1,1}& w_{1,2}  & \cdots  & w_{1,N_{h-1}} \\ 
   w_{2,1}& w_{2,2}  & \cdots & w_{2,N_{h-1}} \\ 
   \vdots & \ddots   & \ddots  & \vdots \\ 
   w_{N_h,1}& w_{N_h,2} & \cdots  & w_{N_h,N_{h-1}}
  \end{bmatrix}
\end{equation*}
and the bias vector in the $h^{th}$ layer is represented as $b_h\in \mathbb{R}^{N_h}$, is understood to be a measure of a neuron's potential activity.
\begin{equation*}
    b_h=
    \begin{bmatrix} 
    
\vspace{.2cm} b^1_h\\ \vspace{.2cm}
b^2_h\\ \vspace{.2cm} 
b^3_h\\ \vspace{.2cm}
\vdots \\ 
b^{N_h}_h
\end{bmatrix}
\end{equation*}

On the above-generated results, an application of a nonlinear activation $\sigma$ function is performed.  Sigmoid, tanh, or ReLU are some of the frequently utilized activation functions in PINNs,
and the operations can be condensed as,
\begin{itemize}
  \item[] \textbf{Input Layer :} $G^1(x)=x\in \mathbb{R}^N$
  \item[] \textbf{Hidden Layer :} $G^h(x)=\sigma(W_hG^{h-1}(x)+b_h)\in \mathbb{R}^{N_h}$
  \item[] \textbf{Output Layer :} $G^H(x)=W_HG^{H-1}(x)+b_H\in \mathbb{R}^{N_H}$ 
\end{itemize}

\subsection{Automatic Differentiation}
For computing the solution of DE using PINNs, the derivative of the output of NN with respect to the input data is considered. Since the NN approximates the solution of DE with a smooth activation function, the output is differentiable. To compute the derivatives, there are four methods: manual differentiation, symbolic differentiation, numerical differentiation, and automatic differentiation (AD). In the case of manual differentiation, the process is not automated, whereas, for symbolic and numerical differentiation, it becomes a cumbersome task to compute derivatives for complex problems. Citing these shortcomings, AD is preferred over other methods. Additionally, it doesn’t have issues like floating point precision error or memory intensity like the other methods and can accurately determine the derivative at a given position which also serves as other motivating factors. When dealing with floating point values, they can be expressed as exact expressions with no approximation error. In AD, a function is decomposed into differentiable sub-functions, and intermediate variables hold the data in the function at different points. Rather than calculating a closed-form expression or estimating the derivative considering the neighboring points, AD takes advantage of this feature, and the chain rule is utilized to combine derivatives of smaller sub-functions through forward and backward modes to determine the numerical value of the derivative. So in PINNs, this differentiation method is employed to compute the partial derivatives of the NN output with respect to the input features, and the computed derivatives are substituted in initial \& boundary conditions, and the governing DE equation. Further, the mean square error of the residual of each of the components in a DE is calculated for the minimization of the loss function of PINNs.

\subsection{Formulating the loss function }

Consider a parameterized PDE expressed in general form,

\begin{equation*}
    \begin{split}
       f(x,t,\frac{\partial y}{\partial x},\frac{\partial y}{\partial t},...; \Psi)&= 0 \hspace{1cm}x\in \Omega,~~t \in [0,T]\\
       y(x,t_0)&=h(x) \hspace{0.5cm} x\in \Omega\\
       y(x,t)&=g(t)\hspace{0.5cm} ~x\in \partial\Omega,~~t \in [0,T]
    \end{split}
\end{equation*}
defined on the domain $\Omega \in \mathbb{R}^N$ with boundary $\partial \Omega$. Where $x=(x_1,x_2,\cdots,x_N) \in  \mathbb{R}^N$ represent the spatial coordinate, $t$ is the time, $f$ is the function describing the problem with differential operators and parameter $\Psi$. Here, $y(x,t)$ is the solution of PDE with initial condition $h(x)$ and boundary condition $g(t)$ which can be Dirichlet, Neumann, Robin or periodic boundary conditions.

With the universal approximation capability of NN, a surrogate solution of $y(x,t)$ is constructed which is represented by $\hat{y}(x,t;\theta)$, where $\theta$ represent the set of weights and bias vectors in a neural network.
$$y(x,t)\approx \hat{y}(x,t;\theta)$$

In order to minimize the divergence between the predicted and the desired solutions, a loss function is constructed based on the governing equation as well as initial and boundary conditions. Hence, AD is used to compute the partial derivative of $\hat{y}(x,t;\theta)$ with regard to input data used in the PDE, which also frames the loss function. Thus, the loss function can be expressed as the weighted summation of $L_2$ norm of the residual of PDE, boundary condition, and initial condition.  The loss function thus formulated is optimized using gradient-based optimizers mostly Adam and the limited-memory Broyden-Fletcher-Goldfarb-Shanno (LBFGS) algorithm, to determine the optimal parameter $\Theta$ that lowers the computational error.

$$\mathcal{L}(\Theta)=w_f\mathcal{L}_f(\theta)+w_{ic}\mathcal{L}_{ic}(\theta)+w_{bc}\mathcal{L}_{bc}(\theta)$$
where
\begin{equation*}
    \begin{split}
      \mathcal{L}_f(\theta)&=\frac{1}{N_f}\sum_{i=1}^{N_f}\left |\left|f(x,t,\frac{\partial \hat{y}}{\partial x},\frac{\partial \hat{y}}{\partial t},...; \Psi)\right|\right|^2_2,\\
      \mathcal{L}_{ic}(\theta)&=\frac{1}{N_{ic}}\sum_{i=1}^{N_{ic}}\left |\left|  \hat{y}(x,t_0)-h(x) \right|\right|^2_2\\
      \mathcal{L}_{bc}(\theta)&=\frac{1}{N_{bc}}\sum_{i=1}^{N_{bc}}\left |\left|  \hat{y}(x,t)-g(t) \right|\right|^2_2
    \end{split}
\end{equation*}
Here, $N_f$ is the set of collocation points, $N_{ic}$ is the set of points that satisfy the initial condition, $N_{bc}$ is the set of points that satisfy the boundary condition and $w_f$, $w_{ic}$ and $w_{bc}$ are the weights.

\section{PINNs for solving ODEs}\label{sec2}
Solving ordinary differential equations (ODEs) has been a complex task over time as mathematical approaches often turn out to be complex as the complexity of the equation increases. Thus with the advent of deep learning, classical DL approaches have been employed in solving ODE projecting them as an ideal approach for generating solutions in a computationally efficient manner. In such instances, the NNs are trained on datasets with known properties in a supervised manner for individuals or groups of similar equations. The knowledge acquired by the NN is further employed to approximate solutions to problems with similar contexts. 

Compared to mathematical approaches, DL-based approaches present a wide set of advantages when attempting to solve ODEs. First and foremost, solutions generated through these approaches are highly accurate irrespective of the complexity of the mathematical approaches involved in solution generation. Secondly, boundary condition and dimensionality are two factors deciding the efficacy of mathematical approaches. However, DL approaches are agnostic to both factors. Finally, data having stochastic distribution or noise also can be easily solved using these methods. Currently, some of the most widely employed deep learning techniques for solving ordinary differential equations are as follows:
\begin{enumerate}
 \item \textbf{Neural Ordinary Differential Equation (Neural ODEs) :} Certain neural networks can be specifically designed to solve a particular set of ODEs in a given scenario. They mainly function by generating a mapping to identify the relationship between the contents of the ODE and their respective solutions. Such NNs are referred to as Neural ODEs~\cite{NEURIPS2018_69386f6b}. Considering the popularity of other approaches, Neural ODE-based approaches are yet to gain traction.
 \item \textbf{Physics-informed neural networks (PINNs):} In the context of ODE, PINNs also work on similar principles as Neural ODEs by utilizing specific NN to solve ODEs. However, PINNs incorporate physical knowledge into the system such that the loss function can be fine-tuned further thus optimizing the solution identification as well.
 \item \textbf{Generative adversarial networks (GANs):} The usage of GANs for solving ODEs does not take a direct approach. Unlike other models, the main motto of employing GANs here is to synthetically generate training data for the NN~\cite{NEURIPS2020_3c8f9a17}. Such data are usually added to the datasets to improve the diversity and thus the accuracy as well.
\end{enumerate}
Developing DL-based approaches for solving ODE is an emerging research domain. However, a considerable volume of literature can be traced in certain methods compared to others. 
We will be limiting our discussion to PINN-related approaches only as the other methods are out of scope for this study.

One of the baseline articles triggering the research community to investigate the applicability of deep learning techniques for solving ODE was presented by Maede et al.~\cite{MEADE199419}. The authors employ one of the basic forms of feed-forward neural networks to identify approximate solutions to nonlinear ODEs. In contrast to classical NN-based approaches, the system does not require to undergo the training phase to perform the approximation. The neural network employs a piece-wise linear map to act as an activation function thus making the storage linear. Even though several articles have claimed it as their research finding, Meade's article was one of the first to identify the relationship between the number of hidden layers of the NN and their impact on the $L_2$ norm of the approximation error.  Other NN parameters such as input data, bias, and weights and their role in generating such networks were also discussed in the article. 
As an extension to \cite{MEADE199419}, Maede proposed using another feed-forward NN-based approach to solve linear ODE in article~\cite{MEADE19941}. The proposed methodology functions based on a hard limit transfer function which makes the system highly linear in storage and function with less processing time. The authors identify that the $L_2$ norm of the overall error generated during approximation is found to be decreasing quadratically as the depth of the hidden layers increases.

Article \cite{mall2013regression} introduces another classical learning-based approach for solving ODE i.e. via an unsupervised variant of the regression-based algorithms. Similar to PINNs, the proposed methodology needs the ODE to have strict initial and boundary conditions and also employs an error back propagation methodology as the former. The backpropagation ensures that both the error function and the model parameters are optimized without employing any specific optimization techniques. 
The authors also make an observation that the depth of hidden layers of the network can be determined considering the degree of polynomial regression 
and the weights are also initialized based on the same regression approach. A similar NN-based approach for solving ODE is proposed by Ezadi et al.~\cite{ezadi} in their article. Compared to the existing approaches, the authors employ a combination of modified NN combined with specific optimization approaches to attain the desired results. The training points for the system are selected over a specific area of open intervals only which reduces the computational errors considerably. In the case of points that do not belong to the aforementioned intervals, they are converted to similar intervals prior to being trained over the network. Additionally, the use of hyperbolic secant as an activation function enhances the accuracy of the modified NN.

The concept of approximating solutions for ODE using another type of neural network, the wavelet neural network (WNN) was presented in their article by Tan et al.~\cite{tan2019solving}. In the proposed NN, the model employs two specifically
designed activation functions, Gaussian wavelet, and Mexican
Hat activation function applied on ODEs with initial conditions only.  
Unlike existing approaches, the authors change the mathematical representation of the problem of solving ODEs from a constraint optimization problem to an unconstrained one strictly following the initial conditions of the equations. Following this, to minimize the unsupervised error, the momentum back propagation approach is employed where the system only attempts to adjust the weights from the hidden to the output layer leaving the other layers untouched. The authors also validate the working of the proposed model against existing ANN approaches.

Even though solving ODE using PINNs has been presented in several existing literature, the depth of the studies is minimal failing to present much findings. However, Nascimento et al.~\cite{NASCIMENTO2020103996} present one of the initial comprehensive discussions on the usage of PINNs for solving ODE. The article gives more emphasis on an implementation perspective using classical Python frameworks. The authors justify the framework selection by stating the ease of availability of neural network models and the inbuilt automatic differentiation features which is a major building block of PINNs. Here, rather than giving more concentration on physical concepts, the authors make use of the data-driven kernel as an easier way to converge the model training. Thus, the hybrid network generated possesses features from both physical concepts as well as that of data-driven kernels. 

Another article on using DL-based techniques for solving ODE was presented by Li et al.\cite{li2021solving} which can be considered as one of the motivating factors behind the development of PINNs. The article proposed a novel methodology identifying numerical solutions for ODE with the aid of improved ANNs. The initial phase of the network was to calculate the approximate solution of the particular ODE followed by loss minimization. The loss function employed in the network was a joint loss function which was formulated by combining several error calculation functions helping to identify errors occurring at a wide set of sample points. Additionally, the network parameters were reformulated based on the findings generated from the Levenberg-Marquardt (RLM) algorithm. The major advantage of the proposed network was its capability to achieve high levels of accuracy and faster convergence compared to the current state of the art. Such a finding has surely motivated the research community to explore better models suited for individual differential equations which have played a major role in the standardization of PINNs.

A similar article on PINNs working on recurrent neural networks (RNN) and data-driven kernels has been proposed by Viana et al.~\cite{VIANA2021106458}. However, the discussion is more centered around the physical aspects, and the authors mainly propose a methodology to estimate missing physical components in NNs. Initially, the RNN is represented as a directed graph making it easier to identify and isolate physics-informed kernels and those that are not. Such a representation makes it easier to evaluate the missing physics at various levels and make changes accordingly. A major advantage of such a technique is that even hidden nodes can be easily adjusted which tends to be a very tedious task in normal cases. The experimental studies incorporating the findings validate the NN's capability to reduce the system loss in a computationally efficient manner. An approach based on parameterization is proposed by Peres et al.\cite{belbuteperes2021hyperpinn} in their article where the proposal of a novel parameterization-based PINN framework referred to as hyperPINNs is made. The proposed NN is capable of solving any given differential equation from the parametric configuration with the aid of hypernetworks. The authors showcase the working of the network on both PDE and ODE given that the differential equation should be maintaining a smaller size irrespective of the size of the parametric space.

 In contrast to the majority of the research on PINNs which is centered around the penalizing approach, Sidharth et al.~\cite{nand2023physicsinformed} utilize the reparameterization approach in their article. The prime motto of the article is to lower the approximation error via reparameterization. Thus, the authors initially estimate the approximation errors which further determines the extent of reparameterization required. For this purpose, the PINNs are subjected to a comparison test where the network undergoes testing with and without reparameterization using the approximation error which gives the precise idea of further reparameterization required. The experimental results demonstrate the validity of the proposed model for fine-tuning approximation error management in complex ODEs.
 \begin{table*}[!hbt] 
 \caption{Summarised table of PINNs used for solving solving ordinary differential equations}\label{table2}%
 \scriptsize
\begin{tabular}{ccll}

\hline
Ref                      & Context                                                                                                  & \multicolumn{1}{c}{Methodology}                                                                                                                                 & \multicolumn{1}{c}{Inferences Derived}                                      
\\ \hline
\cite{MEADE199419}    & \begin{tabular}[c]{@{}c@{}} Nonlinear ODE\end{tabular} & \begin{tabular}[c]{@{}l@{}}A basic feed-forward NN using piece-wise \\linear map activation function is proposed.
\end{tabular}                                         & \begin{tabular}[c]{@{}l@{}}The activation function ensures linear storage and the \\depth of the model is inversely proportional to $L_2$ norm \\of approximation error.\end{tabular}                                          \\ \hline
\cite{MEADE19941}    & \begin{tabular}[c]{@{}c@{}} Nonlinear ODE\end{tabular} & \begin{tabular}[c]{@{}l@{}}A feed-forward NN employing hard limit\\ transfer function is proposed.
\end{tabular}                                         & \begin{tabular}[c]{@{}l@{}}The hard limit transfer function makes the storage linear \\also optimizing the computational time. Increases depth \\of the network result in a quadratic decrease in the $L_2$\\ norm of approximation error.
\end{tabular}                                          \\ \hline

\cite{mall2013regression}    & \begin{tabular}[c]{@{}c@{}} First order ODE \end{tabular} & \begin{tabular}[c]{@{}l@{}}The architecture employs an unsupervised\\ variant of the classical regression-based\\ model.
\end{tabular}                                         & \begin{tabular}[c]{@{}l@{}}The backpropagation method employed in the model \\ensures optimization of the error function and model \\parameters even in the absence of dedicated \\optimizers. However, the model is highly dependent\\ on the initial set of weights to perform well.
\end{tabular}   \\ \hline

\cite{ezadi}    & \begin{tabular}[c]{@{}c@{}} First order ODE \end{tabular} & \begin{tabular}[c]{@{}l@{}}A modified NN with domain-specific \\optimization functions is proposed.
\end{tabular}                                         & \begin{tabular}[c]{@{}l@{}}The combination of reduction in the area of training \\points combined with employing the secant activation \\function reduces computational errors and improves\\ model accuracy.
\end{tabular}   \\ \hline

\cite{tan2019solving}    & \begin{tabular}[c]{@{}c@{}} Linear \& non-homogeneous\\ODE\end{tabular} & \begin{tabular}[c]{@{}l@{}}A wavelet NN combined with customized \\Gaussian wavelet and Mexican-Hat \\activation function is proposed.
\end{tabular}                                         & \begin{tabular}[c]{@{}l@{}} Employing momentum back-propagation approach to \\change hidden layer weights alone yields compatible \\results with lesser computational cost.

\end{tabular}   \\ \hline
\cite{NASCIMENTO2020103996}    & \begin{tabular}[c]{@{}c@{}} First \& Second order\\ODE\end{tabular} & \begin{tabular}[c]{@{}l@{}}The architecture employs an unsupervised \\variant of the classical regression-based model.
\end{tabular}                                         & \begin{tabular}[c]{@{}l@{}} The combination of physical concepts with more \\emphasis on data-driven kernels results in easier \\model convergence.
\end{tabular}   \\ \hline
\cite{li2021solving}    & \begin{tabular}[c]{@{}c@{}} Second-order nonlinear\\ ODE \end{tabular} & \begin{tabular}[c]{@{}l@{}}An improved ANN employing joint \\loss function is proposed.
\end{tabular}                                         & \begin{tabular}[c]{@{}l@{}} Customised models and tailored loss functions deliver\\ better results compared to a generalized model.
\end{tabular}   \\ \hline

\cite{VIANA2021106458}    & \begin{tabular}[c]{@{}c@{}} Standard ODE\end{tabular} & \begin{tabular}[c]{@{}l@{}}An RNN-based PINNs integrated with\\ data-driven kernels is proposed.
\end{tabular}                                         & \begin{tabular}[c]{@{}l@{}} A graph-based representation of the RNN ensures easier\\ identification of both hidden nodes and physics-informed\\ kernels.
\end{tabular}   \\ \hline

\cite{belbuteperes2021hyperpinn}    & \begin{tabular}[c]{@{}c@{}} Standard ODE\end{tabular} & \begin{tabular}[c]{@{}l@{}}A
novel parameterization-based PINNs\\ framework is proposed.
\end{tabular}                                         & \begin{tabular}[c]{@{}l@{}}Hypernetworks facilitate solving PDEs \& ODEs from \\parametric configuration alone given their size is small.
\end{tabular}   \\ \hline

\cite{nand2023physicsinformed}    & \begin{tabular}[c]{@{}c@{}} Standard ODE\end{tabular} & \begin{tabular}[c]{@{}l@{}}A PINNs model employing \\reparameterisation is proposed.
\end{tabular}                                         & \begin{tabular}[c]{@{}l@{}} The iterative reparameterization approach based on \\changing approximation errors fine-tunes the model\\ performance.
\end{tabular}   \\ \hline

\cite{baty2023solving}    & \begin{tabular}[c]{@{}c@{}} Nonlinear ODE\end{tabular} & \begin{tabular}[c]{@{}l@{}} A standard PINNs framework is proposed
\end{tabular}                                         & \begin{tabular}[c]{@{}l@{}}PINNs perform well for PDE with weak linearity.\\ For highly nonlinear PDEs, prior knowledge\\ of data is required.
\end{tabular}   \\ \hline

\cite{baty2023solving2}    & \begin{tabular}[c]{@{}c@{}} Stiff Ordinary \\differential equations\end{tabular} & \begin{tabular}[c]{@{}l@{}}A customised PINN model is proposed.
\end{tabular}                                         & \begin{tabular}[c]{@{}l@{}} The model requires prior knowledge of the problem for\\ generating better approximation. Also, the physical term \\of the loss function has a positive impact on the model \\accuracy for stiff equations.
\end{tabular}   \\ \hline

\end{tabular}
\end{table*}
An article on similar grounds with an investigatory approach is presented by Baty et al.~\cite{baty2023solving}. The authors attempt to present the transition of DL techniques employed for solving ODE all the way from conventional NNs to PINNs. Various factors involved in designing a PINN are explained in detail such as the formulation of the loss function, the role of physical concepts, and the optimization approaches as well. The performance of the network is tested against various ordinary differential equations and validated against classical integration approaches. The authors identified the major advantage of PINNs to be its capability to generate competent results against any of the currently employed techniques with a fractional volume of data compared to the latter given that the problems have weak non-linearity in nature. However, with highly nonlinear problems, PINNs fail to deliver good results in normal conditions and require a priori knowledge of the training data over some level of integration interval to cover up the performance. Additionally, the authors have also critically reviewed the rationale behind incorporating physical knowledge in optimizing PINNs.

The authors have also made an attempt to extend the discussion by presenting an article specifically designed for solving stiff ODE using PINNs~\cite{baty2023solving2}. Often, stiffness is a property that makes the task of solving ODE highly complex for numerical methods. Even with vanilla PINNs, the system fails to perform well thus validating the necessity of the work. The problem often lies in cases with a lack of sufficient training data or large integration intervals. The authors emphasize improving several aspects of the classical PINNs to make them capable of solving stiff ODEs given that the problem has initial conditions and the network has a priori knowledge of the training data. Some of the major findings of the authors include the positively proportional relationship between the physical term in the loss function and the performance accuracy, especially in the context of problems concerning energy conservation tasks. Also, a reformulation of the loss function is necessitated and the possibility of employing an incremental time interval for reducing the residual of ODE is investigated. Finally, the findings are validated on advection-diffusion equations for solving boundary value problems.

A brief summary of the articles discussed in Section~\ref{sec2} on solving ODE is presented in Table~\ref{table2} with their respective methodologies and other features explained in detail.

\section{PINNs for solving PDEs}\label{sec3}

Partial differential equations are a subgroup of DEs involving multiple independent variables, a function dependent on the variables, and a partial derivative of the function. PDEs abide by several physical laws making it an ideal instrument for the investigation of underlying patterns and trends in various natural phenomenons. Additionally, PDEs are also employed for analyzing several engineering problems related to fluid dynamics, thermal dynamics, and so on. Thus, streamlining solving PDEs still holds high research value considering its versatility and applicability in several domains. However, a major hurdle related to the study of PDE is the inability to generate efficient analytical solutions till date. Even though several well-established numerical approaches are available, researchers have been in constant search of newer methodologies stating the complexity of the existing approaches.

Considering the developments in computational mathematics, DL-based methods have found their way into solving PDEs as well. Even basic neural networks have high levels of approximation capabilities when working on nonlinear functions. In contrast to the numerical approaches, inherent NN features such as statistical learning capabilities and optimization methods make it possible to solve a wide array of PDEs irrespective of their linearity and dimensionality. Even though simpler PDEs still employ statistical approaches stating their merits, such DL-based approaches have been widely employed for solving complex PDEs in recent times. Faster and more accurate approximation capabilities, generalizability, and several other features have also played a prominent role in accelerating the adaptation of DL approaches. Following the trend, specific frameworks have been developed for solving specific PDE-based problems over time such as PINNs.

One of the initial articles on solving PDEs using PINNs was proposed by Zhang et al.\cite{10.1137/19M1260141}. The article presented one of the trivial versions of PINNs for solving stochastic PDEs. The model employs dynamic orthogonal and borthogonal methods for accommodating the stochasticity of data. The loss function was formulated using these constraints combined with the decomposed values of the stochastic PDE to fine-tune the approximations. Since the proposal was in the initial stages of the technology, the models generated were not as efficient and accurate as in its current state. However, inspired by these works, several attempts to solve similar equations and families of equation has been made using PINNs. Recently, Karlbauer et al.\cite{pmlr-v162-karlbauer22a} proposed a variant of the classical PINNs termed compositional finite PINNs or FINNs specifically for solving spatiotemporal advection-diffusion procedures. The proposed architecture combines the features of classical ANN with physical and structural knowledge gained from the numerical simulation of PDEs. The model has proved its applicability for a variety of 1-D and 2-D PDEs such as the Burgers equation, diffusion equation, and so on. The FINN framework sets itself apart from both conventional ML approaches as well as standard physics-informed approaches mainly due to its generalization ability and robustness toward problems even from unknown distributions.

As a majority of DL approaches mainly employed grid-based models, employing grid-free models for solving PDEs was identified to be an effective solution for certain complex problems. Thus, Pu and Feng \cite{e24081106} proposed a PINN architecture based on grid-free learning for solving coupled Stokes–Darcy equations with Bever–Joseph–Saffman interface conditions. The usual DL approaches employ a grid-based learning schema meaning that the input data is represented in the form of grids during training. Such approaches are usually computationally complex taking more time to generate inferences and do not function well for complex input data. However, with grid-free approaches, complex problems can be easily solved using a fraction of time as compared to the former. In the case of Stokes–Darcy equations, the presence of small parameters and certain interface discontinuity proves it to be harder for classical PINNs to approximate the solutions. Thus the proposed model utilizes several strategies such as an updated weighted loss function, an adaptive local function, and so on. Even though the model excels in approximating solutions, the true extent of the grid-free nature of the network has not been harnessed yet as it is evident from the convergence speed of the network which still remains a questionable factor.

In article~\cite{Xia_2023}, authors propose a spectrally adapted PINN architecture for solving analytically intractable PDEs. Intractable PDEs are notoriously difficult to solve via numerical approaches as the equations often contain variables defined in unbounded domains. However, the equations also hold the key to solving several engineering and biomedical problems. The proposed s-PINNs employ a hybrid approach combining features of classical PINNs along with adaptive spectral principles. s-PINNs functions by identifying the physical disparity among various classes involved such as time and space components and using the inference to interpret spatial data more effectively by building a basis function. Unlike standard PINNs, s-PINNs can solve unbounded and complex PDEs without the need for a high volume of data but suffer from increasing network dimensionality. 

Another variant of PINNs for solving second-order parametric light wave equations is proposed by Sun et al.\cite{e25040674}. From the existing literature, the authors identified the computational complexity involved in the standard automatic differentiation approaches employed for solving PDEs of higher order. Inspired by the findings, the authors presented the gPINN architecture built using an updated neural network structure termed the second-order neural network structure. The central difference of the PDEs are considered while formulating the network structure. The gPINN architecture is comprised of an adaptive activation function along with a gradient enhancement strategy for accelerating the convergence of the system. To compensate for the computational overhead generated by the enhanced gradient, an additional deep-mixed residual model was incorporated into the architecture. The authors present extensive experimental results on various forward and inverse problems where the model performed exceptionally well for estimating and reconstructing missing parameters in the latter group.

Wight and Zhao~\cite{wight2020solving} proposed an enhanced PINNs architecture for solving PDEs used for interfacial domain problems, namely the family of phase field equations such as Allen-Cahn and Cahn-Hilliard equations. Even though PINNs perform exceptionally for a large number of PDEs, phase-filed equations are an exception. The authors investigate the effect of various sampling techniques on the approximation power of the PINN model by interpreting both the spatial and temporal content present in the data with equal importance. Listing down the benefits of differential approaches, the authors combine the classical time and adaptive sampling approach to formulate the time adaptive sampling approach citing its capability of attaining better accuracy and convergence in the particular use case. The proposed approach adopts a methodology similar to mini-batch processing where every individual network in the framework focuses on smaller problem domains extracted from the bigger one. A major issue observed in the system was the inability of the model to be robust against various errors occurring in much smaller time fragments. This resulted in the propagation of such errors occurring in a small fragment to the whole integral over time. Thus, the applicability of the method becomes limited to problems with very large time domains only where it becomes hard for the model to deviate from the preset bounds.

A major issue faced in solving PDE using PINNs is the increasing network complexity with the complexity of the input equation. Even though several methods for identifying derivatives from highly complex equations have been proposed, most of them fall short in terms of accuracy. Thus, article \cite{mukhametzhanov2022high} proposes a highly accurate novel higher-order differentiation method for time-dependent PDEs for a set of spatial points. The derivative thus obtained acts as an intermediate that can be applied to several data-driven procedures for generating final solutions. The authors provide an experimental perspective of the study by showcasing the capability of the proposed approach to incorporate further data points dynamically and its positive impact on the solutions generated. The strategy had also been tested and validated on the  Burgers’, Allen-Cahn, and Schrodinger equations. In a similar context, a short article discussing the applicability of PINNs for solving Burgers equations was presented in \cite{10141811}. The authors propose a custom PINN architecture with a 9-layer structure with integrated tanh activation functions. Rather than concentrating on the actual PINN architecture itself, the majority of the study revolves around identifying the effects of the Adam optimizer on the loss function when solving the specific PDE. 

Fang et al.\cite{8946607} also proposed a novel PINN framework for solving time-dependent constrained surface PDEs. The existing PINNs frameworks employed a direct approach where the PDEs would be solved directly. For normal PDEs, this approach usually works as it is often observed that when the loss function of a PINN reaches zero, the residue of the equation also goes to zero which directly points to the ability of the network to work efficiently on the training data. However, this observation does not fit well for surface PDEs. Thus, the proposed model utilizes a set of points instead of a partition or a mesh as used in classical frameworks. The formulation of such point clouds simplifies the system considerably. Even though comparative studies amongst various numerical approaches testify to the functionality of the model, the authors fail to present a theoretical proof of error. The authors also state the applicability of the model for other problems related to moving surfaces and other time-dependent problems as possible future extensions. Another article on solving time-dependent PDE is presented by Wu et al. \cite{10.1007/978-3-031-08754-7_45}. The authors present a novel PINN architecture based on the classical RNN and discrete cosine transform functionalities combined with the standard MLP framework. The DCT encodes the spatial content and the RNN interprets the temporal values. Combined together, they generate a latent content grid to be fed to the MLP model. The grid fine-tunes the PINNs network when dealing with spatiotemporal data. Compared to the existing models, the proposed model excels on the Taylor-Green vortex problem with Navier-Stokes equations

Even though state-of-the-art PINN approaches have simplified solving nonlinear PDEs considerably, PDEs with discontinuous solutions still remain a fairly unexplored problem. In the article \cite{lv2021hybrid}, the authors investigate discrete time PINN and propose a hybrid PINN architecture by integrating a discontinuity indicator for solving nonlinear PDEs. The role of the indicator is to isolate the smooth and nonsmooth regions of the input data. Following that, the proposed hPINNs utilize the classical automatic differentiation for computing various differential operators of the equation in smooth regions and an updated weighted essentially non-oscillatory scheme is employed to identify nearby discontinuity patterns. The hPINNs architecture outperforms the classical variants on various  Burger equations when approximating discontinuous solutions even over wider gaps.

Pakravan et al.\cite{PAKRAVAN2021110414} presented a novel framework, Blended inverse-PDE networks (BiPDE)  for identifying missing frames in inverse PDE problems. The framework attempts to uncover the hidden parameters involved in the \textbf{retro} using domain-specific knowledge and physical components in a finite number of iterations. In contrast to the existing approaches which contain their structural knowledge of the loss function alone, in BiPDE, the classical neural networks take up the role of universal functions estimators along with a PDE solver layer for strictly governing the physical components. Thus, the computation is streamlined as the system can focus mainly on performing the prescribed task alone. The proposed model is validated on several high dimensional and inverse PDE problems such as Poisson equations in varying spatial domains, Burgers equations, and so on.

Another group of equations where PINNs struggle to perform well is PDEs with a point source simply due to the singularity brought into the domain by the Dirac delta function allied with the equation. Thus, Huang et al. \cite{huanguniversal} proposes a universal approach for solving such equations by combining a set of three different techniques. Initially, the authors propose reformulating the Dirac delta function as a continuous probability density function to get rid of the singularity factor. Following that, the training loss of the system is balanced by employing a lower bound constrained uncertainty weighting algorithm. Finally, a multi-scale DNN is utilized for the actual network itself to improve the model convergence. Approximations performed on physical problems validate the functionality of the proposed model based on its performance efficacy, approximation capabilities, and versatility.

In article \cite{9829676}, a cartesian coordinate integrated approach to solve PDEs using PINNs is proposed. The authors perform a comparative analysis of results generated using PINNs with and without integration of coordinates and are evaluated for two different instances, systems with periodicity and systems that are void of it. The studies reveal that systems using cartesian coordinates showcase better accuracy during approximations. Extended analysis performed using other coordinate systems such as polar coordinates and symmetric cylindrical coordinate systems also testifies the efficacy of employing Cartesian approach in PINNs frameworks.

\begin{table}[!htb] 
 \caption{Summarised table of PINNs used for solving partial differential equations}\label{table3}%
 \scriptsize
\begin{tabular}{ccll}
\hline
Ref                      & Context                                                                                                  & \multicolumn{1}{c}{Methodology}                                                                                                                                 & \multicolumn{1}{c}{Inferences Derived}                                                                                                                      \\ \hline

\cite{pmlr-v162-karlbauer22a} & \begin{tabular}[c]{@{}c@{}}Spatiotemporal advection-diffusion \\procedures  \end{tabular}                                                                                                     & \begin{tabular}[c]{@{}l@{}}The framework combines features of ANN \\with physical and structural inferences from\\ numerical simulation.\end{tabular} & \begin{tabular}[c]{@{}l@{}}The hybrid approach enhances generalizability, \\accuracy and robustness even outside boundary \\conditions for various 1D and 2D PDEs.\end{tabular} \\ \hline

\cite{e24081106} & \begin{tabular}[c]{@{}c@{}}Stokes–Darcy equations with \\Bever–Joseph–Saffman \\interface condition    \end{tabular}                                                                                                 & \begin{tabular}[c]{@{}l@{}}A grid-free DL method inspired PINNs \\ framework is proposed\end{tabular} & \begin{tabular}[c]{@{}l@{}} The presence of smaller parameters combined with \\interface discontinuity prevents the framework from \\gaining full utility of grid-free approach even after \\employing techniques such as weighted loss function, \\local adaptive activation function and so on.\end{tabular} \\ \hline

\cite{Xia_2023} & Analytical Intractable PDEs                                                                                                      & \begin{tabular}[c]{@{}l@{}} A hybrid PINN framewok combining features \\of classical PINNs and adaptive spectral \\methods is proposed.\end{tabular} & \begin{tabular}[c]{@{}l@{}}Eventhough use of spectral concept improves accuracy,\\ convergence \& applicability to unbounded domain \\problems, the model suffers from increased NN \\evaluation cost as problem complexity increases.\end{tabular} \\ \hline

\cite{e25040674} &  \begin{tabular}[c]{@{}c@{}}Second-order parametric light \\wave equations     \end{tabular}

& \begin{tabular}[c]{@{}l@{}}A second-order neural network with adaptive\\ activation and a gradient enhancement
\\strategy is proposed.\end{tabular} & \begin{tabular}[c]{@{}l@{}}Introducing deep-mixed residual model\\ compensates for the increase in computation cost \\\end{tabular} \\ \hline

\cite{wight2020solving}    & \begin{tabular}[c]{@{}c@{}}Interfacial domain problems \\ such as phase field equations\end{tabular} & \begin{tabular}[c]{@{}l@{}}A combination of novel time adaptive sampling \\approach combined with domain fragmentation \\and mini batch processing is proposed.\end{tabular}                                         & \begin{tabular}[c]{@{}l@{}}The proposed mini match processing facilitates\\ propagation of errors from mini-batches to whole\\ domain making the system less robust.\end{tabular}                                          \\ \hline

\cite{mukhametzhanov2022high} & \begin{tabular}[c]{@{}l@{}}Nonlinear time dependent \\PDEs \end{tabular}                                                                                                     & \begin{tabular}[c]{@{}l@{}}A higher-order differentiation method for \\generating
set of spatial points is proposed.\end{tabular} & \begin{tabular}[c]{@{}l@{}}The derivates generated via high order differentiation\\ method can be fed into various data-driven procedures \\for generating solutions.\end{tabular} \\ \hline

\cite{8946607} & \begin{tabular}[c]{@{}c@{}} Time dependant eliptical \\PDEs on 3D surfaces  \end{tabular}                                                                                                        & \begin{tabular}[c]{@{}l@{}}A grid free point cloud based PINNs \\framework is proposed.\end{tabular} & \begin{tabular}[c]{@{}l@{}}The usage of point clouds and their respective \\normal values ensures lesser errors and streamlines\\ approximations for surface PDEs.\end{tabular} \\ \hline
\cite{10.1007/978-3-031-08754-7_45} & \begin{tabular}[c]{@{}c@{}} Time-dependent \\partial differential equations   \end{tabular}                                                                                                        & \begin{tabular}[c]{@{}l@{}}
A PINN framework functioning on the latent\\ representation of spatiotemporal \\components is proposed. \end{tabular} & \begin{tabular}[c]{@{}l@{}}Using DST for encoding spatial frequencies \& RNNs \\for temporal features generates a compressed\\ representation optimizing functioning of PINNs.\end{tabular} \\ \hline

\cite{lv2021hybrid} & \begin{tabular}[c]{@{}c@{}} Non linear PDEs with \\discontinuous solutions   \end{tabular}                                                                                                        & \begin{tabular}[c]{@{}l@{}}A PINN framework integrated with a\\ novel discontuinty indicator module \\is proposed.\end{tabular} & \begin{tabular}[c]{@{}l@{}}Classifying local solutions based on their continuity \\and employing AD and WENO scheme for smooth \\and discontinuous fragments ensures better \\approximations even for larger time intervals. \end{tabular} \\ \hline

\cite{PAKRAVAN2021110414} & \begin{tabular}[c]{@{}c@{}} Unknown field identification \\for inverse PDE problems   \end{tabular}                                                                                                  & \begin{tabular}[c]{@{}l@{}}A blended approach leveraging expressibility \\of DNNs as universal estimators \& knowledge \\from numerical methods embedded into \\semantic encoder layers.\end{tabular} & \begin{tabular}[c]{@{}l@{}}Hard-coded PDE solvers enable the network to be \\aware of underlying physics resulting in a \\streamlined process solely concentrating on hidden \\field identification only. \end{tabular} \\ \hline

\cite{huanguniversal} & \begin{tabular}[c]{@{}c@{}} PDEs with a point source \\expressed as a Dirac \\delta function   \end{tabular}                                                                                                        & \begin{tabular}[c]{@{}l@{}}
A PINN framework with a universal solution \\based on three novel methods for solving\\ Dirac delta function is proposed. \end{tabular}
 & \begin{tabular}[c]{@{}l@{}}
 Converting the Dirac delta function into a probability \\distributed function combined with a balancing loss\\ function via maintaining a contained lower bound\\ through uncertainly weighted algorithms enables\\ the multi-scale DNN to perform better. 
 \end{tabular} \\ \hline

\cite{9829676} & \begin{tabular}[c]{@{}c@{}} Standard Partial differential\\ equations\end{tabular}                                                                                   & \begin{tabular}[c]{@{}l@{}}A cartesian coordinate integrated \\ PINNs framework is proposed \end{tabular} & \begin{tabular}[c]{@{}l@{}}Employing cartesian coordinate systems to PINNs \\ensures better approximation capabilities without\\ the need for coordinate transformation as in \\traditional analytical approaches.\end{tabular}  \\ \hline

\cite{REN2022114399} & \begin{tabular}[c]{@{}c@{}} Spatiotemporal partial \\differential equations   \end{tabular}                                                                                                        & \begin{tabular}[c]{@{}l@{}} An unsupervised PINNs framework based\\ on convolutional-recurrent learning \\scheme is proposed.\end{tabular} & \begin{tabular}[c]{@{}l@{}}The unsupervised learning scheme facilitates better \\extraction of spatial data at low dimensions \& \\temporal
evolution learning.\end{tabular} \\ \hline

\cite{9533606} &  \begin{tabular}[c]{@{}c@{}}Standard Partial differential\\ equations \end{tabular}                                                                                                & \begin{tabular}[c]{@{}l@{}}A novel learning scheme employing multi-task \\learning techniques combined with
uncertainty\\weighting loss and gradients surgery is proposed.\end{tabular} & \begin{tabular}[c]{@{}l@{}}Learning on a combination of shared representations \\ \& adversarially generated high loss samples of similar\\ PDEs helps the model to improve generalization.\end{tabular} \\ \hline

\cite{10.1007/978-3-031-36024-4_42} & \begin{tabular}[c]{@{}c@{}} Standard Partial differential\\ equations      \end{tabular}                                                                                                 & \begin{tabular}[c]{@{}l@{}}A hierarchical architecture based PINNs\\ frameworks is proposed.\end{tabular} & \begin{tabular}[c]{@{}l@{}} Multi-level training ensures efficient capture of \\various frequency band components at respective \\levels ensuring faster convergence. \end{tabular} \\ \hline

\cite{BIHLO2022111024} & \begin{tabular}[c]{@{}c@{}}Shallow water equations on a \\sphere with applicability to \\meteorological systems  \end{tabular}                                                                                               & \begin{tabular}[c]{@{}l@{}} A multi-model approach for solving equations \\with longer time intervals is proposed.

\end{tabular} & \begin{tabular}[c]{@{}l@{}} Fragmenting larger time intervals and employing an \\NN on each fragment yields better results. Encoding \\boundary conditions into the NN can help manage\\ boundary loss efficiently.\end{tabular} \\ \hline

\cite{wassing2022parametric}    & \begin{tabular}[c]{@{}c@{}}Motion flow prediction via Euler's \\ \& Navier Stokes equations\end{tabular} & \begin{tabular}[c]{@{}l@{}}An artificial dissipation integrated training \\approach for PINNs is proposed.\end{tabular}                                         & \begin{tabular}[c]{@{}l@{}}Dissipation term facilitates filtering of data further \\ aiding in faster\end{tabular}                                          \\ \hline

\cite{wassing2023physicsinformed} & \begin{tabular}[c]{@{}c@{}} Parametric compressible Euler \\equations with applicability to\\fluid dynamics problems \end{tabular} 
&\begin{tabular}[c]{@{}l@{}}An adaptive artificial viscosity reduction \\method integrated PINN model is proposed.\end{tabular} & \begin{tabular}[c]{@{}l@{}}
Introducing penalty term combined with equation   \\parameters fed as input dimensions facilitate\\ learning at continuous parametric space for viscosity\\factor estimation.  \end{tabular}\\ \hline 
 
\end{tabular}
\end{table}
The majority of current literature on PINNs for solving PDE is centered on supervised approaches. However, these approaches rely heavily on fully connected NNs and hyperparameters, sometimes failing to accommodate low-dimensional spatiotemporal data. Motivated by these issues, Ren et al.\cite{REN2022114399} proposed a novel unsupervised variant of PINNs referred to as physics-informed convolutional-recurrent network or PhyCRNet. The system architecture comprises an encoder-decoder convolutional long short-term memory which is capable of performing feature extraction even for low dimensional spatial and temporal features. In contrast to the loss function used in supervised models, the loss function in PhyCRNet is formulated as the combined aggregate of PDE residual fragments.  Additionally, the framework also utilized autoregressive and residual connects to synthesize time marching. The authors investigate the model performance on different nonlinear PDEs such as Burgers equations, FitzHugh Nagumo reaction–diffusion equations, and so on. Compared to the existing approaches, the proposed model attains high levels of accuracy and generalizability for unlabelled data.  The presence of hard-coded I/BCs also acts as an optimization approach during training which further aids in streamlining the system towards faster convergence.

Even though several proposals were presented using grid-free approaches to solve PDEs, the system was still facing performance deterioration once the domain nonlinearity had been introduced into the system. Thus, in an attempt to increase robustness and attain better generalizability, Thanasutives et al.\cite{9533606} presented a novel multi-task learning approach combined with uncertainty-weighted loss and gradient surgery for PDE-related problems. The proposed multi-task learning scheme exploits the possibility of knowledge sharing over a set of similar PDEs by utilizing the PDE parametric coefficients. The procedure is orchestrated by cross-stitch modules and ensures that the initial PDE gains better generalization ability post-knowledge sharing. The working scheme is similar to the one adopted in general domain adaptability problems where the system attempts to fine-tune the original use case itself from the knowledge obtained from the inferred case. The authors also propose employing adversarial training for stimulating high-loss samples similar to the training data. The introduction of adversarial training combined with multi-task learning helps the system focus more on high-loss regions which are usually challenging to interpret for normal approaches thus enhancing the system's performance over a wide range of test data including highly complex stochastic PDEs.

A major issue associated with solving PDEs using computational approaches is the inability of the models to interpret both low and high-frequency components of the input data with the same level of expertise. Thus, lower-frequency data gets processed much more quickly compared to the high-frequency components which automatically decreases the overall convergence time. Taking this into account, Han and Lee~\cite{10.1007/978-3-031-36024-4_42} proposed a hierarchical learning framework specifically suited for solving PDEs. The system works in a hierarchical format meaning that any neural network layers added to the existing network learn from the residual data of the predecessor's approximation. The hierarchical architecture is implemented using an ML model followed by a Fourier feature embedded network. The initial network aids in capturing the low-frequency values of the solution with ease followed by the feature embedding layer imposing the range of scales of the solution. Although the architecture has been tested over several existing linear and nonlinear PDEs, certain aspects have still been left unaddressed. One such major issue has been associated with the inability to properly justify the relativity between network complexity and characteristic scales when representing the function.

Apart from solving classical mathematical problems, applying PINNs for PDEs satisfying various real-world use cases has also found wide applicability. One such article is proposed by Bihlo et al.\cite{BIHLO2022111024}  where the authors envisaged the applicability of PINNs from a meteorological context by solving the shallow-water equations on the sphere in their article. In contrast to the existing PINN with a single model, the author proposes employing a multi-model approach to increase the range of the neural networks to equations with longer time intervals. Splitting the time interval of the equation into smaller exclusive fragments and doing training on each to generate inferences is also proposed. However, for the boundary conditions, the final prediction of the predecessor acts as the initial condition for the successive networks. In addition to this, the introduction of such boundary conditions for training the NN results in a system with nullified boundary value loss conditions which further simplifies the optimization procedures. Instead, the author's proposal is to encode the boundary conditions to the respective network itself. Articles \cite{wassing2022parametric} and \cite{wassing2023physicsinformed} are also of a similar nature but from the perspective of fluid-related problems.

In article \cite{wassing2022parametric}, the applicability of PINNs to estimate the motion of fluid via Eulers and Navier Stokes equations has been presented. Even though the existing computational methods were highly advanced, they still had several shortcomings along with a very limited scope for further optimization. The authors identify that the employment of PINNs could easily overcome classical issues faced when using first principle-based solvers. The PINNs employed in the networks utilize artiﬁcial dissipation technique during the model training. The dissipation term allows the filtering of unphysical data from the result and also caters to dealing with numerical viscosity in fluid-based problems when the system nears convergence. Wassing et al.~\cite{wassing2023physicsinformed} also presented their article from a similar context. Here, the connection to fluid dynamics problems is established via stationary compressible Euler equations. The proposed PINN model is built on adaptive artificial viscosity reduction methods for subsonic and supersonic data in multiple dimensions. The network is structured to accommodate the parameters of the equations as additional input dimensions thus enabling the model to function in continuous parametric space. Since the model is developed for fluid dynamic problems, the major aim is to predict the viscosity factor locally. It is made possible with the addition of a penalty term during the model training. The authors also experimentally validate their claims comparing their findings with vanilla PINNs. As the discussion on such articles can't be limited to a few, an extensive study has been presented in Section~\ref{sec6}.

 The summary of the study for PDE is presented in Table~\ref{table3} with a detailed notion of their working methodology and findings from each article

\section{PINNs for solving fractional differential equations}\label{sec4}

Compared to PINN-based approaches for solving PDE and ODES, fractional PDE is a less investigated domain with very few articles published proposing PINN-based solutions. However, we will be attempting to present a comprehensive summary of the available literature. A major building block of the classical PINNs architecture is the automatic differentiation feature. Even though only minimal changes are required when shifting from ODE to PDE, the classical AD fails to perform well in the case of fractional PDE. The reason is the inability of the methodology to calculate fractional derivatives. The basic principle of the AD is derived from the standard chain rule which becomes inapplicable in fractional calculus, thus presenting challenges in the basic fundamental architecture itself. In general,  a fractional variant of PINNs referred to as fPINNs has been proposed to tackle these issues. However, researchers have been working around the problem proposing enhancements to the fPINNs architecture and also figuring out newer methods to solve the problem.

Inspired by the limitations of PINNs for solving fractional PDEs, the initial proposal of fPINNs was made by Pang et al. in their article \cite{doi:10.1137/18M1229845}. Since AD becomes useless in the context of fractional equations, the authors devised a novel hybrid approach to generate the residual value in the loss function. The proposed approach utilizes the standard automatic differentiation for values of integer order and then uses numerical discretization for fractional data.  For experimental validations, the article mainly dealt with the advection-diffusion equation having space-time-fractional components. Even though considered as a benchmark for solving fractions PDEs, the article has limited applicability. However, with updations, it can easily solve other equations belonging to differential, integral, and integro-differential classes as well. Following the trend, Yan et al.\cite{yan2023laplacefpinns} have proposed an extended version of the classical fPINNs with additional Laplacian-based features for solving forward and inverse fractional diffusion equations. The approach is termed Laplace-based fPINNs or Laplace-fPINNs in short. The combination of the Laplacian transform along with the numerical inverse Laplacian transform removes the need for using auxiliary points which reduces the complexity of the problem considerably. The proposed approach excels for inverse problems with high dimensionality and also has the scope to be extended to other subdiffusion equations as well.

Another attempt at solving time-fractional phase field equations has been made by Wang et al. in their article \cite{wang2022fractional}. The proposed framework combines the features of fPINNs along with spectral collocation methods to approximate the results. Unlike classical fPINNs, the proposed model has large representation capabilities due to the presence of spectral collocation. The proposed spectral collocation approach fragments the space direction and the fractional backward difference formula approximates the derivative of the time fractional data. It also reduces the volume of discrete fractional operators employed which also plays a crucial role in improving the system's accuracy. The model functions efficiently for both forward and inverse problems with similar neural network architecture. However, the network structure needs to be tuned according to the use case to generate a trade-off between generalization and learning capabilities. A sampling-based variant of the fPINN architecture is proposed by Guo et al.\cite{GUO2022115523} for solving backward and forward fractional PDEs. The approach utilizes the Monte Carlo quadrature-based approximation methods to generate the derivatives from the fractional part of data and generate loss function with unbiased physical constraints, thus termed Monte Carlo fPINNs or MC-fPINNs. In other existing variations of fPINNs,  a substantial increase in the number of auxiliary points occurs when attempting to fragment the fractional derivative, especially when the dimensionality of the input increases. However, with this approach, the model ensures that the auxiliary points remain low irrespective of the complexity of the input equation which makes the framework highly streamlined. In general, the model excels against high dimensional fractional equations such as fractional Laplacian equations, fractional diffusion equations, and so on.

\begin{table}[!h]
\caption{Summarised table of PINNs used for solving fractional differential equations}\label{table4}%
\scriptsize
\begin{tabular}{ccll}

\hline
Ref                      & Context                                                                                                  & \multicolumn{1}{c}{Methodology}                                                                                                                                 & \multicolumn{1}{c}{Inferences Derived}                                      
\\ \hline
\cite{doi:10.1137/18M1229845}

& \begin{tabular}[c]{@{}c@{}} Standard fractional PDE\end{tabular} & \begin{tabular}[c]{@{}l@{}}A standard fPINNs model with\\ updated loss function is proposed.
\end{tabular}                                         & \begin{tabular}[c]{@{}l@{}}The combination of automatic differential for integer order \\values and numerical discretization for fractional order overcome \\the shortcomings of standard automatic differentiation procedure.\end{tabular}                                          \\ \hline
\cite{yan2023laplacefpinns}    & \begin{tabular}[c]{@{}c@{}} Forward and inverse fractional\\ diffusion equations\end{tabular} & \begin{tabular}[c]{@{}l@{}}A customised fPINNs framework \\with Laplacian features is \\proposed.
\end{tabular}                                         & \begin{tabular}[c]{@{}l@{}} Combination of Laplacian transform and numerical inverse \\Laplacian transform removes the need for using auxiliary points\\ thus reducing problem complexity and making it ideal for\\ high-dimensional data.
\end{tabular}                                          \\ \hline

\cite{wang2022fractional}    & \begin{tabular}[c]{@{}c@{}} Time-fractional phase\\ field equations\end{tabular} & \begin{tabular}[c]{@{}l@{}}
 An updated fPINNs framework \\integrated with spectral collocation \\methods is proposed. \end{tabular}                                    & \begin{tabular}[c]{@{}l@{}}
Combining the spectral collocation approach for fragmenting \\spatial components and the fractional backward difference\\ formula for approximating time fractional components reduces\\ the volume of discrete fractional operators thus improving \\system accuracy.

\end{tabular}   \\ \hline

\cite{WANG202364}    & \begin{tabular}[c]{@{}c@{}} Variable Order space\\fractional PDE\end{tabular} & \begin{tabular}[c]{@{}l@{}} Two PINNs frameworks with power \\series expansion and dual network \\architecture is proposed.
\end{tabular}                                         & \begin{tabular}[c]{@{}l@{}}

Power serious expansion yields a better result for forward \\problems as it can generate approximations without the need \\for fragmenting fractional values. For inverse problems, a dual\\ network architecture combined with power series coefficient is \\used to determine fractional order and generate approximations.

\end{tabular}   \\ \hline

\cite{GUO2022115523}    & \begin{tabular}[c]{@{}c@{}} Backward and forward \\fractional PDE\end{tabular} & \begin{tabular}[c]{@{}l@{}} A PINNs model integrated \\with Monte Carlo approximation\\ is proposed.
\end{tabular}                                         & \begin{tabular}[c]{@{}l@{}} Custom formulated unbiased loss function combined with the \\reduction in auxiliary points ensuring better results even for\\ high dimensional data.

\end{tabular}   \\ \hline
\cite{HAJIMOHAMMADI2021111530}    & \begin{tabular}[c]{@{}c@{}} Nonlinear ODE\end{tabular} & \begin{tabular}[c]{@{}l@{}}  A framework combining classical\\ DNN and Chebyschev collaboration \\approach is proposed.
\end{tabular}                                         & \begin{tabular}[c]{@{}l@{}} The activation function leverages the orthogonal bias to generate \\approximation in a computationally efficient manner.
\end{tabular}   \\ \hline
\cite{ma2023biorthogonal}    & \begin{tabular}[c]{@{}c@{}} Time-dependent stochastic\\ fractional PDE\end{tabular} & \begin{tabular}[c]{@{}l@{}}
A PINNs framework integrated with\\ biorthoganalism is presented.
\end{tabular}                                         & \begin{tabular}[c]{@{}l@{}} Combination of grid-based discretization and weakly formulated \\loss function generated better results for inverse and forwards \\fractional problems.
\end{tabular}   \\ \hline
\cite{REN2023126890}    & \begin{tabular}[c]{@{}c@{}} Fractional PDE with \\oscillations \& singular\\ perturbations\end{tabular} & \begin{tabular}[c]{@{}l@{}} Fractional PINNs employing an\\ improved layer based weight \\initialisation strategy
\end{tabular}                                         & \begin{tabular}[c]{@{}l@{}} Combination of cubic polynomials, adaptive methods and \\enhanced physical knowledge yields better results for oscillatory \\and perturbed data.
\end{tabular}   \\ \hline

\end{tabular}
\end{table}

Another approach for solving fractional PDEs, especially the time-dependent stochastic fractional PDEs was proposed by Ma et al.\cite{ma2023biorthogonal}. The authors propose combining the concepts of biorthoganalism for accommodating the stochastic nature of the equation and combining it with the classical PINNs thus formulating a newer method termed bi-orthogonal fPINN or BO-fPINN. In place of automatic differentiation, the model employs grid-based discretization to generate the fractional derivatives. Similarly, the loss function is also substituted with a weakly formulated one to surpass the inherent issues faced in BO methods. The major advantage of the BO-fPINN framework is its ability to perform with similar expertise for both inverse and forward problems using the same network formulations. The model functions efficiently for a wide set of stochastic fractional variants of the classical PDEs such as 1D nonlinear reaction-diffusion, 2D reaction-diffusion, and so on.

Another customized version of the fPINNs is proposed by Ren et al.\cite{REN2023126890} designed for differential equations with certain levels of oscillations or perturbations. The framework is termed ifPINNs as it employs an improved approach for weight initialization between layers based on a nonlinear strategy as compared to a linear approach in PINNs. The model employs a combination of cubic polynomial and adaptive methods depending on the use case for oscillatory and perturbed data. The updated strategy ensures better utilization of the physical information thus ensuring better results for both cases. The framework also has the capability to function equivalently to classical PINNs or fPINNs by changing certain parameters which makes them equally functional for fractional PDEs as well. However, the degenerated ifPINNs yield better results than their initial counterparts in the majority of the cases. The proposed framework has superior fitting ability over a wide range of equations as it uses a higher volume of hyperparameters which leaves the question of hyperparameter optimization and selection as a future research direction of ifPINNs.

Apart from fPINNs, certain other variants of PINNs have also been proposed for solving fractions PDES. One such PINN framework is proposed by Wang et al.\cite{WANG202364} for solving a specific type of fractional PDE belonging to the variable-order (VO) space-fractional class. The authors attempt to solve both forward and backward problems for VO space fractional advection-diffusion problems using the same framework but with slight variations. For forward problems, power serious expansion combined with classical PINNs is utilized. Once the approximate solution is generated from the power series expansions, the fractional derivatives are not further subjected to discretization and can be directed to the PINN algorithm. Similarly for inverse problems, a dual network architecture inspired by principles of parallelism is utilised to determine the order of the fractional operator. The power series coefficient is utilized here as well for generating the coefficients along with another function to determine the order of the fractional operator. The strategy has better generalisation capabilities and can be easily adapted to data with higher dimensions without compromising accuracy and convergence speed.

Haji Mohammadi et al.\cite{HAJIMOHAMMADI2021111530} has also proposed another DL-based framework inspired from spectral methods for solving fractions DEs. The method is termed as Fractional Chebyschev Deep Neural Network or FCDNN which is a combination of a classical DNN framework with the Chebyschev collaboration approach. The framework is termed after the optimization employed, the Chebyshev orthogonal polynomial approach. The activation function approximates the solution with the aid of orthogonal bases reducing the computations considerably. Post approximations,  the solutions pass through the derivative and integral nodes where the residual function is formulated. Apart from these, the classical components of DNNs and PINNs components such as optimizers, activation function, and the automatic differentiation for finding the derivatives are respectively employed.

Even though numerous articles can be identified in this particular problem, we are forced to limit our discussion to a selected few due to the limited scope. Table~\ref{table4} summarizes the articles we have investigated and provides
an elite comparison based upon their methodology and features.

\section{Variants of PINNs}\label{sec5}
In this section, we attempt to discuss some of the major variants of PINNs based on their working methodology and area of applicability. Even though several other frameworks can be identified, we limit our discussion to some of the articles that possess a certain level of architectural novelty or research contribution rather than a mere extension of existing literature.
\begin{enumerate}
\item \textbf{Conservative PINNs :} The conservative PINNs (cPINNs) is a novel framework proposed by Jagtap et al.\cite{JAGTAP2020113028} that employs non-linear conservative laws for solving discrete domains. In classical PINNs, the domains are further fragmented into subdomains, and the cPINNs function at these subdomain levels. The conservative property of the framework is ensured by fluxing a rigid continuity among all the subdomains. Additionally, the concept of the average solution is also integrated at the convergence point of two such subdomains where the result generated at two different NNs are averaged. For NN architecture, even though the proposed article employs a local adaptive activation function, the framework provides a certain degree of leniency when it comes to choosing the activation function and other parameters such as optimization approach, the volume of residual points, NN depth, and so on. As the model leverages the power of domain decomposition combined with domain-based NN functionalities, it becomes synonymous with parallel computation systems where each fragment can be considered as a single computational entity.
\item \textbf{Extended physics-informed neural networks :} Jagtap et al.\cite{JAGTAP2020113028}  further extended their work on physics-informed ML by proposing a generalized PINN architecture combining features of both the classical PINNs and cPINNs termed extended physics-informed neural networks (XPINNs) in artcile\cite{jagtap2021extended}. The XPINNs framework employs spatio-temporal domain decomposition techniques and is mainly applied for solving nonlinear PDEs on arbitrary complex-geometry domains. In contrast to standard PINNs, the proposed model has better representation and parallelization capabilities. It follows a similar pattern to cPINNs by using different NNs for individual fragments post-decomposition. However, the major novelty of the model lies in its capability to adapt to any PDE, unlike its predecessors. Additionally, it also offers the feature of arbitrary domain decomposition thus facilitating spatial and temporal parallelization which ultimately leads to streamlined training. When it comes to NN selection for each fragment, XPINN framework facilitates using customized NNs with domain-specific parameters such as NN depth, optimization approach, and so on. The XPINNs also offer parallelization capabilities as cPINNs since it is highly generalizable in terms of both domain applicability and decomposition.
\item \textbf{Augmented PINNs}: A further improved version of the XPINNs is also proposed under the title  Augmented PINNs or APINNs~\cite{HU2023107183} where a soft and trainable domain decomposition approach is employed along with flexible parametric selection for improved performance. The framework architecture utilizes a trainable gateway network to mimic the decomposition capabilities of XPINNs which plays a remarkable role in generating better fragments. Additionally, training data utilization among various fragments is also facilitated in the model ensuring generalizability at the partition level as well. The combined utilization of partial parameter sharing and weighted averaging among subnetworks governed by the partition-of-unity feature of the proposed gateway network delivers the final output of the model. The overall performance of APINNs turns out to be far superior to the XPINN models in terms of domain fragmentation, subnetwork management, generalizability, and so on. Currently, optimized versions of APINNs are also being proposed under different titles~\cite{GUAN2023112360}.
    
    \item \textbf{Fractional PINNs :} The general proposals of PINNs were mainly centered around solving PDE and ODE problems. However, another major variant of DEs, fraction differential equations was less emphasized in the literature of PINNs. One of the major components of PINNs, namely the automatic differentiation function was so well customized for integer order DEs that they failed to accommodate the fractional components present in the equation completely. Thus a dedicated variant of PINNs for solving fractional DEs were proposed, termed fPINNs.  The fPINNs framework employs a hybrid approach by disintegrating the integer order values and fractional order values and generating the residual loss function for both using automatic differentiation and numerical differentiation respectively. This approach helps to overcome the issues associated with the standard PINNs for dealing with fractional operators. Unlike other variants discussed here, fPINNs is an area of high research interest with extensive research being presented at prestigious venues. We limit our discussion here to a minimal introduction only as a detailed study of the topic is already presented in Section~\ref{sec4}.
\item  \textbf{Distributed PINNs :} Complying with the global change happening in AI framework development from centralized to distributed paradigms, articles proposing distributed variants of the PINN framework also have been popping up. One such article based on distributed PINNs for solving PDE is proposed by Dwivedi al.\cite{dwivedi2019distributed}. The authors refer to the model as DPINNs and the major aim of the approach is to present a comprehensive evaluation of the classical PINNs based on the representation capabilities of the models. Similar to \cite{HU2023107183}, DPINNs also employ a domain fragmentation approach and run a classical PINN model in each fragment. This makes the problems simpler and capacitates models with lesser complexity to deliver satisfactory results. Once training is completed, the loss function simply aggregates the individual losses from each fragmented domain eventually yielding better results than the classical centralized version of PINNs. Such a method also ensures better representation when implemented on nonlinear PDEs such as Burgers equation and Navier-Stokes equations. The physical components present in the neural networks automatically take up the role of regularizers further boosting the representation capabilities of the model.
    
\item \textbf{Interval and fuzzy PINNs}:   Solving PDEs with uncertain spatial and temporal contents is a cumbersome task for any DL model. Even though finite element methods have succeeded in solving these equations, their applicability is still limited to use cases with known correlation information only. Thus, customized PINN frameworks involving both fuzzy and integral concepts have been proposed by Fuhg et al.~\cite{FUHG2022103240} termed fuzzy PINNs (fPINNs) and intervalPINNs (iPINNs) respectively. The proposed fPINNs functions based on the convex fuzzy set principles and the iPINNs are built combining two separate NNs, one for approximating primary output and the second one for estimating the allied input data that points to the PDE solutions. In contrast to existing models, the system requires no a priori knowledge of the input data and still manages to retain all the inherent advantages of PINNs such as grid-free architecture, expertise on both inverse and forward problems, and so on. The model is generally applicable for identifying bounded solutions from any PDEs when the spatial and temporal parameters cannot be properly defined due to uncertain fields.

\end{enumerate}
\section{Application of PINNs}\label{sec6}

In this section, we will be discussing various real-life applications of PINNs and also stating their impact on the respective domains as well. A short comparison of the existing methods and motivation for shifting to PINNs is stated from the implementation perspective for applications related to medical systems, power systems, fluid mechanics and so on.

\subsection{Medical Domain :} \label{biomedical}    
 With the increasing applications of DL technologies in the medical domain, huge research has been happening in the field of DL-integrated drug development.  Even though drug-target interaction is one such domain with high volumes of literature with DL techniques, a major issue faced is the inability of the models to generalize and perform under computational constraints.  Thus, Moon et al.\cite{D1SC06946B} proposed a PINN-based solution for enhancing existing DTI approaches termed PIGNet. The model has two different components, namely a PINN network for predicting the pairwise connections between atoms and an affinity term depicting the effect of binding between the protein-ligand compound. The physical component present in the network facilitates the model to generate better inferences behind the chemical interactions taking place inside the delivery location. For attaining further generalization, the authors propose performing data augmentation with other protein ligands having diverse chemistry before being fed to the PINN models. Experimental validations and ablation studies performed testify to the system's performance and generalizability compared to existing approaches.
 
Another application of PINNs in the medical sector was proposed by Sarabian et al.\cite{9740143} for cerebrovascular disease detection via hemodynamic analysis in the brain. The authors proposed employing PINNs to augment clinical data for generating brain hemodynamic patterns at higher resolutions. The current state-of-the-art non-invasive method known as transcranial doppler (TCD) ultrasound has certain limitations with respect to its locations of applicability. Thus the proposed model was built from TCD-based velocity measurement and other conventional methods to estimate various parameters sythetically. The authors also clinically validate the proposed model for the diagnosis of cerebral vasospasm by synthetically generating blood flow data. A similar article on hemodynamic parameter estimation for cardiovascular scenarios is presented by Zhang et al.\cite{ZHANG2023107287}. The authors propose combining principles of fluid dynamics along with PINN models to stimulate flow field datasets. The model when tested on anisotropic data validates the capabilities of PINNs in solving patterns within spatiotemporal data for bodily fluid motions such as blood flow in vessels with the use of accurate personalized models.
    
The application of PINNs in medical systems is not just limited to hemodynamics or similar mechanisms. A framework for cardiac systems is proposed by Buoso et al.\cite{BUOSO2021102066} where a parametric PINNs architecture was proposed for simulating personalized left ventricular biomechanics system. The model draws inferences from high-quality cardiac images of the left ventricle to generate radial bases which are then combined with a neural network to estimate the cardiac mechanisms. The models function in a computationally cheaper and more efficient manner compared to the existing methods such as finite element models incurring only a fraction of error. Unlike existing methods, the proposed approach also facilitates high levels of personalization as the framework draws inferences from individuals' images. Another article on cardiac systems \cite{10.3389/fphy.2020.00042} proposes a PINN-based approach for cardiac activation mapping to identify underlying wave dynamics and also analyze the model's lack of knowledge of the predictions generated. The current state-of-the-art models utilize sparse data point-based interpolation mapping which has several shortcomings. However, anatomic activation maps are one such problem that has high application in several diagnostic procedures such as atrial fibrillation diagnostics. Thus, the PINNs-based approaches have high relevance in these contexts as they provide a streamlined alternative with less susceptibility to errors
\subsection{Power Systems :}

Rather than just limiting to medical applications, the underlying physical components in PINNs also makes them an ideal method for designing various complex system such as power systems. For complex systems, the PINN models use the input data to learn system dynamics by reducing the error between self-generated predictions and the preset physical laws. This makes PINNs function ideally for a variety of tasks in power systems such as estimating system dynamics, flow analysis, estimating parameters, anomaly detection and isolation, power planning, and so on. Several articles have been presented discussing the applicability of PINNs on power systems. We will be presenting a very selected number of articles here as extensive survey articles\cite{9743327} are already available on the topic.

    Apart from the mainstream problem, PINNs have also been used for solving a unique aspect of power systems related to load margin. Murilo E. C. Bento \cite{10098912} had tackled the fairly unexplored problem by proposing a unique PINN-inspired framework termed  Physics-Guided Neural Network or PGNN for estimating the load margin. In contrast to traditional methods that function by solving differential algebraic equations, the PGNN framework uses auxiliary procedures to reconstruct the power flow equations that directly determine the load margins. The model employs a combinatory loss function minimization approach, one for the empirical knowledge part and one for the physical knowledge part.
    
Similarly, a study on the estimation of the dynamic nature of power systems in a real-time manner was proposed by Mohammadian et al.\cite{MOHAMMADIAN2023109551} recently. Stating the inability of the existing approaches to cope with the dynamic nature of the parameters involved in power systems, gradient-enhanced physics-informed neural networks, or GPINNs is proposed which harness the underlying physical components of power systems. The proposed model is capable of generating estimation in real time with far less volume of training data making the whole procedure computationally efficient. Another PINN-inspired framework termed DAE-PINN~\cite{moya2023dae} was proposed exclusively for simulating solution trajectories of nonlinear differential-algebraic equations (DAE). The generated simulations are applicable to general power systems with a certain degree of stiffness. The DAE-PINN  framework is a combination of the time-stepping schemes derived from the Runge-Kutta method and the classical PINNs.  Thus, the NN approximates the hard constraints of the DAE via a penalty-based approach also simulating the DAE over long time horizons.

    \subsection{Fluid Mechanics:}
Along with other domains, fluid mechanics is another area where PINNs have been employed extensively. The majority of the applications in fluid mechanics fit into the Navier Stokes equations (NSE) and this group of equations can be easily approximated using PINN models. Even though standard numerical approaches exist, there is still room left for newer and more efficient methodologies for solving NSE. In their current form, PINNs and their variants are not mature enough to completely replace the existing numerical approaches. However, they can aid in approximating solutions for equations of higher order which usually tends to be a cumbersome task when using numerical approaches. 

One of the initial proposals of PINNs was centered around estimating unknown parameters from NSE from the velocity of fluid flowing across a cylinder~\cite{RAISSI2019686}. Inspired by the proposal, a wide range of flow problems have been analyzed and solved using PINNs related to compressible flows, biomedical flows, and so on. The majority of articles presented in section~\ref{biomedical} on body fluid estimation can be termed as a sub-track of the fluid mechanic applications. Unlike numerical methods, the major advantage of PINNs in fluid mechanical applications is the capability of the same model to function equally on both forward as well as inverse problems. Additionally, PINNs also function well for combining the observations made from flow data and the physical components i.e. the governing equations. The applicability of PINNs for fluid dynamic problems has also been presented extensively. Thus we will be limiting our study to some of the most recent and relevant articles only.

Apart from biological fluid dynamics, applying PINN-based solutions to equations in a different context, i.e. chemical reactor systems is discussed in~\cite{choi2022physics}. The authors formulate a model of a continuous stirred tank reactor (CSTR) with Van de Vusse reaction using PINNs for performing training on the above-mentioned equations. The PINN architecture for the study is inspired by the classical PINNs proposed in~\cite{raissi2019physics} and the authors also propose an updated version of the network with additional features of mini-batch training and updated loss function for solving issues associated with memory errors and model divergence.

Another article for solving inverse supersonic compressible flow problems which are one type of popular equation employed in aerospace vehicular design problems was proposed by Jagtap et.al~\cite{JAGTAP2022111402}. The authors propose fragmenting the domain into smaller subdomains and employing neural networks at each fragment to identify their respective solutions. To this end, the article presented two variants of PINNs, a standard one and an extended one (XPINN) capable of solving such problems. Towards the context of solving multiphysics fluid flow processes, Aliakbari et al.~\cite{ALIAKBARI2022109002} proposed a novel hybrid approach inspired by standard PINNs. As high-fidelity and low-fidelity models have certain shortcomings individually, the authors propose a multi-fidelity approach where the data is generated by low-fidelity CFD methods and PINN models are incorporated into the system to attain better accuracy. As generating high-fidelity data is a computationally expensive task, the model utilizes low-fidelity data and the PINN network runs on a physics-guided initialization scheme which further compensates for the lack of high-fidelity data. The comparative studies' against the existing model not only prove the capability of the model to improve the accuracy on low fidelity but also testify to the ability to attain faster model convergence and system accuracy.

 Even though several other domains can also be identified where PINNs have established themselves such as nuclear reactor parametric estimations~\cite{ANTONELLO20233409}, thermo dynamics~\cite{MASCLANS2023100448}, seismic activity prediction~\cite{ZOUJingBo}\cite{borate2023using}, soil content analyis~\cite{haruzi2023modeling} and so on, we are forced to limit to some of the major applications only to remain under the scope of the study.

\section{Challenges and future directions}\label{sec7}
Extensive research on PINNs and similar physics-informed learning approaches has uncovered several shortcomings associated with the training and other aspects of the technology. Even though the issues associated with respective frameworks may vary, some of the common problems include convergence issues, computational cost, spectral bias, inability to deal with data stiffness, and so on. Apart from these ML-related aspects, certain mathematical areas of PINNs are also in its primary stages. For instance, the majority of the existing literature lacks theoretical or mathematical proofs which makes their applicability a questionable process. Similarly, convergence analysis, error estimation, and so on are other prospective areas of research in PINNs. In this section, we will be discussing some of the major shortcomings of PINNs and the future directions of research in detail.
\begin{enumerate}
    \item \textbf{Optimization approaches and loss functions:} As with any ML approach, optimization plays a major role in PINNs as well. In contrast to classical NNs, PINNs employ a drastically different training methodology which makes the existing optimization approaches unsuitable for PINNs framework. The same issue applies to loss functions as well. Even though several attempts to streamline these concepts have been proposed, they are far from being considered mature or viable proposals. Thus, developing better optimization approaches and mathematically proven loss functions has extensive research scope in the context of PINNs.
    \item \textbf{Learning architecture :} Similar to the classical DL approaches, PINNs also employ neural networks as their backbone. Thus, the scope for further optimization of NN is prevalent in PINNs as it is applicable to any DL framework. The general optimization may be from various aspects related to architecture, activation function, or even hyperparameter tuning. Since classical DL is a more mature domain, several novel techniques and components have been integrated into the DL frameworks such as normalization approaches, various unsupervised and semi-supervised methods, transformer models, and so on depending upon the complexity of the problems and application domains. However, these aspects are yet to be made familiar in the context of PINNs which leaves room for exploration in the future.
    \item \textbf{Dimentionality Problems:} As PDEs are used to represent major concepts of science and engineering with underlying physical components, several equations require high dimensionality to justify the representation of complex characteristics of the application domains. However, solving these sorts of equations is notoriously difficult due to the curse of dimensionality that they are accompanied with. Since NNs possess an inherent capability of dealing with high-dimensional data in their native domains, the integration of physical knowledge in PINNs further opens up newer horizons for effectively solving high-dimensional PDEs and DEs in general. Thus, it remains an open research domain yet to be fully explored as solving high-dimensional PDEs holds the key to many problems in areas related to quantum theory, robotics, and so on.
    \item \textbf{Model Generalisablity:} Currently, one of the major issues associated with PINNs is the lack of generalization capability of the model being developed. Even for a single family of equations, a universal model does not fit the requirements thus needing a personalized model for every use case. This becomes a cumbersome process in several cases where the complexity of developing models exceeds the benefits from the generation of solutions via computational procedures. Similarly, lack of robustness is also another issue closely related to the model generalization ability. Even though current research has progressed to the level where a single PINNs model is capable of solving both forward and inverse problems, an idealistic model is still far from being developed. Thus, ample research scope exists in this area of PINNs as well 
    \item \textbf{Expressive Ability :} In classical DL, the expressive ability of an NN refers to its ability to represent various types of functions. In the context of PINNs, we specifically mean the capability of the model to estimate numerical operators when using the term. In recent times, the expressive ability of PINNs has been highly researched and considerable progress has been made. However, this feature is directly determined by the depth of the NN being employed which increases with respect to the complexity of the equations. Even though the rationale behind the fact still requires better explanations, it can be identified that deeper NNs have better approximation capabilities compared to shallow NNs. Thus, in-depth studies are required to uncover the relation between the depth of NN and their approximation abilities. Also, developing optimized shallow NNs with higher approximation abilities for complex equations will increase the applicability of the models to lightweight systems making it a prospective area of research.

\item \textbf{Convergence Analysis :} Convergence analysis of any NN algorithm is a keen factor in determining the system's efficiency, the same applies to PINNs as well. However, the current research on PINNs is still in its initial stages lacking formal proofs for several functions. Similar to the need for model personalization for individual problems, the same persists with convergence analysis as well. A standard framework is yet to be developed in this respect thus making the problems complicated.

\item \textbf{Estimating Model Error :} Error estimation is another major factor needing further research in PINNs. Unlike classical DL approaches, PINNs models are mainly estimated based on two variants of errors, namely approximation error and generalization error. The approximation error is associated with the capability of the model to approximate the results from the DEs. Optimizing approximation errors helps in generating better models and algorithms. When it comes to generalization errors, the ability of the model to be robust against various scenarios is evaluated. Optimizing generalization errors results in stable algorithms with improved generalization capacities and robustness. Uncovering factors associated with loss function also plays a major part in ensuring optimal utilization of the physical components in the data.

\end{enumerate}


\section{Conclusion}\label{sec8}

This study provided a comprehensive analysis of physics-informed neural networks from various perspectives including methodologies, variants, the strategy employed for various types of equations, areas of applicability, and so on. We also highlighted the relevance of selecting components such as model architecture, loss function, integration of physical constraints, and their effect on problem domains. The study also investigated some of the prevalent research gaps in PINNs and proposed potential research directions for optimizing the technology. These findings include developing better optimizing strategies, personalized loss functions, formulating novel theoretical proofs, and improving generalizability. Finally, this survey article combines various findings related to PINNs bringing diverse aspects under an umbrella that can serve as a point of reference to researchers working in the domain and provides insights into state-of-the-art methods, challenges, and prospective research directions.

\bibliographystyle{unsrtnat}

\bibliography{cas-refs}





\end{document}